\documentclass[aps,pre,twocolumn,superscriptaddress,groupedaddress,nofootinbib,showpacs]{revtex4}

\usepackage[caption=false]{subfig}
\usepackage{amsmath}    
\usepackage{xcolor}
\usepackage{graphicx}   
\usepackage{hyperref}   
\begin{document}

\makeatletter
\def\scheme{\@float{scheme}}
\let\endscheme\endfigure
\makeatother

\title{Asymmetry between Activators and Deactivators in Functional Protein Networks}
\author{Ammar Tareen}
\altaffiliation{Present Address: Simons Center for Quantitative Biology, Cold Spring Harbor Laboratory, Cold Spring Harbor, New York, 11724, USA}
\affiliation{Department of Physics, Clark University, Worcester, Massachusetts 01610}
\author{Ned S. Wingreen}
\email{wingreen@princeton.edu}
\affiliation{Lewis-Sigler Institute for Integrative Genomics, Carl Icahn Laboratory, Washington Road, Princeton, New Jersey 08544}
\author{Ranjan Mukhopadhyay}
\email{ranjan@clarku.edu}
\affiliation{Department of Physics, Clark University, Worcester, Massachusetts 01610}

\date{\today}
\begin{abstract}
Are ``turn-on'' and ``turn-off'' functions in protein-protein interaction networks exact opposites of each other? To answer this question, we implement a minimal model for the evolution of functional protein-interaction networks using a sequence-based mutational algorithm, and apply the model to study neutral drift in networks that yield oscillatory dynamics. We study the roles of activators and deactivators, two core components of oscillatory protein interaction networks, and find a striking asymmetry in the roles of activating and deactivating proteins, where activating proteins tend to be synergistic and deactivating proteins tend to be competitive. 
\end{abstract}

\maketitle


\section{Introduction}

Biological oscillators are ubiquitous ~\cite{1998PhT....51Q..86G, Mori2009, Nakajima414} and they are often quite complex with many interacting components. How has evolution arrived at such complex networks where oscillations arise from interactions among a large number of components? A reasonable hypothesis is that complex bio-oscillators evolved from simpler core oscillatory modules.  
While biological oscillators often involve both genetic and protein components, it has been suggested that in many systems the protein circuit acts as the core oscillator \cite{FERRELL2011874}. With this in mind, we focus here on oscillatory protein networks, where oscillations emerge from the interplay between positive and negative feedback loops. A central biologically motivated question is how do these feedback loops evolve while maintaining the function of the network? 

The regulation of function in protein interaction networks is often achieved by post-translational modifications of component proteins. A common example is phosphorylation and dephosphorylation \cite{doi:10.1021/cr000225s}. In most cases, an activating protein (e.g. a kinase) covalently modifies a target to activate it, and a deactivating protein (e.g. a phosphatase) reverses this change. 
However, despite this symmetry at the molecular level, at the network level activation and deactivation have distinct roles, and so it is natural to ask if there is an asymmetry in the way activators and deactivators are organized in protein networks. 
Recent studies have highlighted such asymmetries experimentally \cite{smoly2017asymmetrically}. In particular, Smoly \textit{et al.} \cite{smoly2017asymmetrically} performed quantitative analyses of large-scale ``omics" datasets from yeast, fly, plant, mouse, and humans and uncovered an asymmetric balance between kinases and phosphatases - each organism contained many different kinases, and these were balanced by a small set of highly abundant phosphatases. Motivated by this study, we ask whether an asymmetry can arise simply from the intrinsically different roles that activators and deactivators play in protein-interaction networks.

\section{Evolutionary Model}

To answer this question, we adopt a physically-based protein-protein interaction model that allows us to map from sequence space to interactions and consequently to network dynamics, in order to study the evolutionary dynamics of oscillatory networks composed of activators and deactivators \cite{PhysRevE.97.040401,PhysRevE.97.020402}. As in the original network model \cite{PhysRevE.97.040401}, we include cooperativity by assuming that activation or deactivation of a target (itself either an activator or a deactivator) requires $h$ independent binding/modification events, with partially modified intermediates being short lived. This yields the following chemical processes for the two classes of proteins,
\begin{gather}
h A^*_i + {A_j}/{D_j} \overset{k_{ij}}{\rightarrow} h A^*_i + {A^*_j}/{D^*_j}, \nonumber
\\
h D^*_i + {A^*_j}/{D^*_j}  \overset{k_{ij}}{\rightarrow} h D^*_i + {A_j}/{D_j},
\end{gather}
where $A$ represents activators, $D$ represents deactivators, and an asterisk indicates the active form. Note that in our model activators and deactivators act both as enzymes and as targets for the action of other enzymes. The corresponding chemical kinetic equations can be approximated as (see \cite{PhysRevE.97.040401} for details):
\begin{gather}
\frac{dA^*_j}{dt} = \sum_{i=1}^{m} k_{ij} [A^*_i]^h [A_j] -  \sum_{i'=1}^{n} k_{i'j}[D^*_{i'}]^h[A^*_j] + \alpha[A_j] - \beta[A^*_j],  \nonumber
\\
\frac{dD^*_j}{dt} = \sum_{i=1}^{m} k_{ij} [A^*_i]^h [D_j] - \sum_{i'=1}^{n}k_{i'j}[D^*_{i'}]^h[D^*_j] + \alpha[D_j] - \beta[D^*_j],
\label{dkstaridpstari} 
\end{gather}
where $m$ and $n$ are the number of distinct types of activators and deactivators respectively, \ $\alpha$ and $\beta$ represent background activation and deactivation rates respectively, and $h$ represents the degree of cooperativity.

Protein-protein interaction strengths are generally determined by amino-acid-residue interactions at specific molecular interfaces. Moreover, it has been estimated that greater than 90\% of protein interaction interfaces are planar with the dominant contribution coming from hydrophobic interactions \cite{Heo4258,doi:10.1021/cr040409x}. For simplicity, we therefore assume each protein possesses a pair of interaction interfaces, an in-face and an out-face, and we associate a binary sequence, $\vec{\sigma}_{\text{in,out}}$, of hydrophobic residues (1s) and hydrophilic residues (0s) to each interface. The interaction strength between a protein (denoted by index $i$) and its target (denoted by index $j$) is determined by the interaction energy $E_{ij} = \epsilon \vec{\sigma}_{\text{out}}^{i} \cdot \vec{\sigma}_{\text{in}}^{j}$ between the out-face of the protein and the in-face of its target. $\epsilon$ represents the effective interaction energy between two hydrophobic residues. (All energies are expressed in units of the thermal energy $k_B T$.) The reaction rate is then given by
\begin{equation}
k_{ij} = {\Big[\frac{k_0}{1 + \exp[-(E_{ij} - E_0)]}\Big]^h},
\label{kil}
\end{equation}
where $E_0$ plays the role of a threshold energy, e.g. accounting for the loss of entropy due to binding. The background activation and deactivation rates are set equal and define the unit of time via $\alpha = \beta = 1$. In our simulations we set $k_0 = 10^4$, $\epsilon = 0.2$, cooperativity $h = 2$, $E_0 = 5$, and we take the length of each sequence representing an interface to be 25. These interaction parameters are chosen to provide a large range for the rate constants $k_{ij}$ as a function of sequence and to keep the background rates small compared to the highest enzymatic rates; cooperativity is introduced to allow oscillations in relatively simple biomolecular networks.

For our evolutionary scheme, we assume a population sufficiently small that each new mutation is either fixed or entirely lost \cite{1958PCPS...54...60M, nowakbib}. We consider only point mutations - namely replacing a randomly chosen hydrophobic residue (1) in the in- or out-face of one enzyme by a hydrophilic residue (0), or vice versa. In this study, mutations are accepted if and only if they satisfy the selection criterion that the network remains oscillatory.

\section{Activation and Deactivation Asymmetry}

\begin{figure}
\includegraphics[scale=1.5]{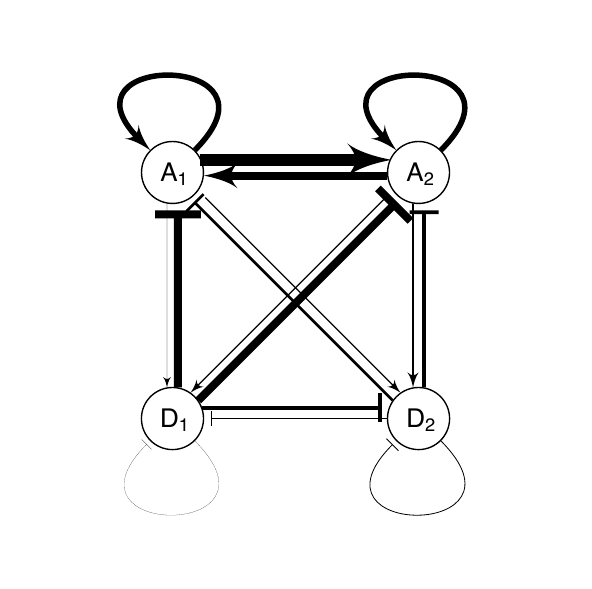}
\caption{Schematic representation of a 2-activator 2-deactivator protein interaction network. Nodes represent proteins while edges represent interactions with other proteins (arrows represent activation while bars represent deactivation). Edge width is proportional to the strength of interaction. This network was obtained by duplication of a 1-activator 1-deactivator network followed by $10^4$ accepted mutations.}
\label{networkt500}
\end{figure}

\begin{figure}
\includegraphics[scale=0.3]{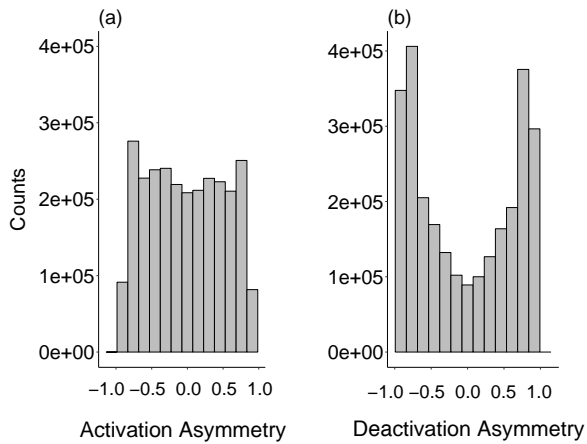}
\caption{Distributions of (a) Activation Asymmetry and (b) Deactivation Asymmetry, constructed from 3 million accepted mutations for a 2-activator 2-deactivator network.}
\label{Asymmetry_Statistics}
\end{figure}

We start with a 1-activator 1-deactivator oscillatory network (the smallest network in our model that can generate oscillations) and duplicate it to generate a 2-activator 2-deactivator network (2A-2D), which is subsequently allowed to evolve  \cite{10.1371/journal.pone.0039052}. Fig. \ref{networkt500} shows a schematic of such a 4-node network after $10^4$ accepted mutations. The widths of the edges are proportional to the interaction strengths and already suggest an asymmetry between activators and deactivators. As the network evolves, does this asymmetry become more prevalent, fluctuate, or disappear, and why? 
To answer these questions, we evolve the system over long evolutionary times (millions of accepted mutations), and consider the distributions of activation and deactivation asymmetries in rate constants, defined as.
\begin{equation}
\begin{aligned}
\textrm{Activation Asymmetry} = \frac{k_{\textrm{A}_{1} \rightarrow \textrm{D}_{1} } - k_{\textrm{A}_{2} \rightarrow \textrm{D}_{1} }}{k_{\textrm{A}_{1} \rightarrow \textrm{D}_{1} } + k_{\textrm{A}_{2} \rightarrow \textrm{D}_{1} }}, \\
\textrm{Deactivation Asymmetry} = \frac{k_{\textrm{D}_{1} \rightarrow \textrm{A}_{1} } - k_{\textrm{D}_{2} \rightarrow \textrm{A}_{1} }}{k_{\textrm{D}_{1} \rightarrow \textrm{A}_{1} } + k_{\textrm{D}_{2} \rightarrow \textrm{A}_{1} }}.
\end{aligned}
\end{equation}  
Activation Asymmetry is a measure of the difference in activation of a single deactivator by two activating proteins. Similarly, Deactivation Asymmetry measures the difference in deactivation of a single activator by two deactivating proteins. We expect these variables to be distributed differently if there is indeed an asymmetry present between activators and deactivators. Fig. \ref{Asymmetry_Statistics}  shows distributions of activation and deactivation asymmetries for a 2A-2D network, constructed from 2 million accepted mutations. 
We indeed find a striking difference in the two distributions, with clear bimodality in the distribution of the Deactivation Asymmetry in contrast to the broad distribution in Activation Asymmetry. 

\begin{figure}
\includegraphics[scale=0.32]{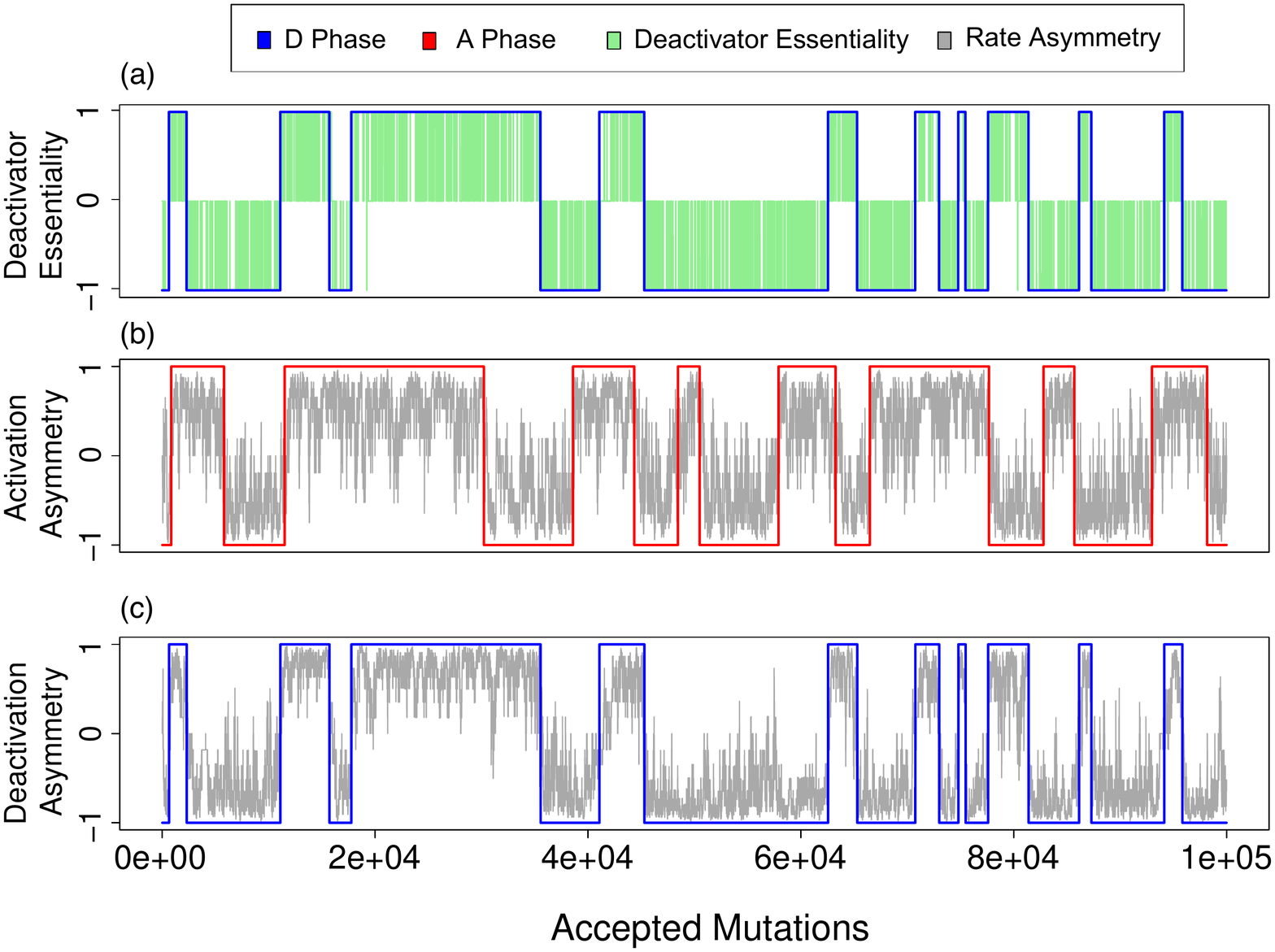}
\includegraphics[scale=0.34]{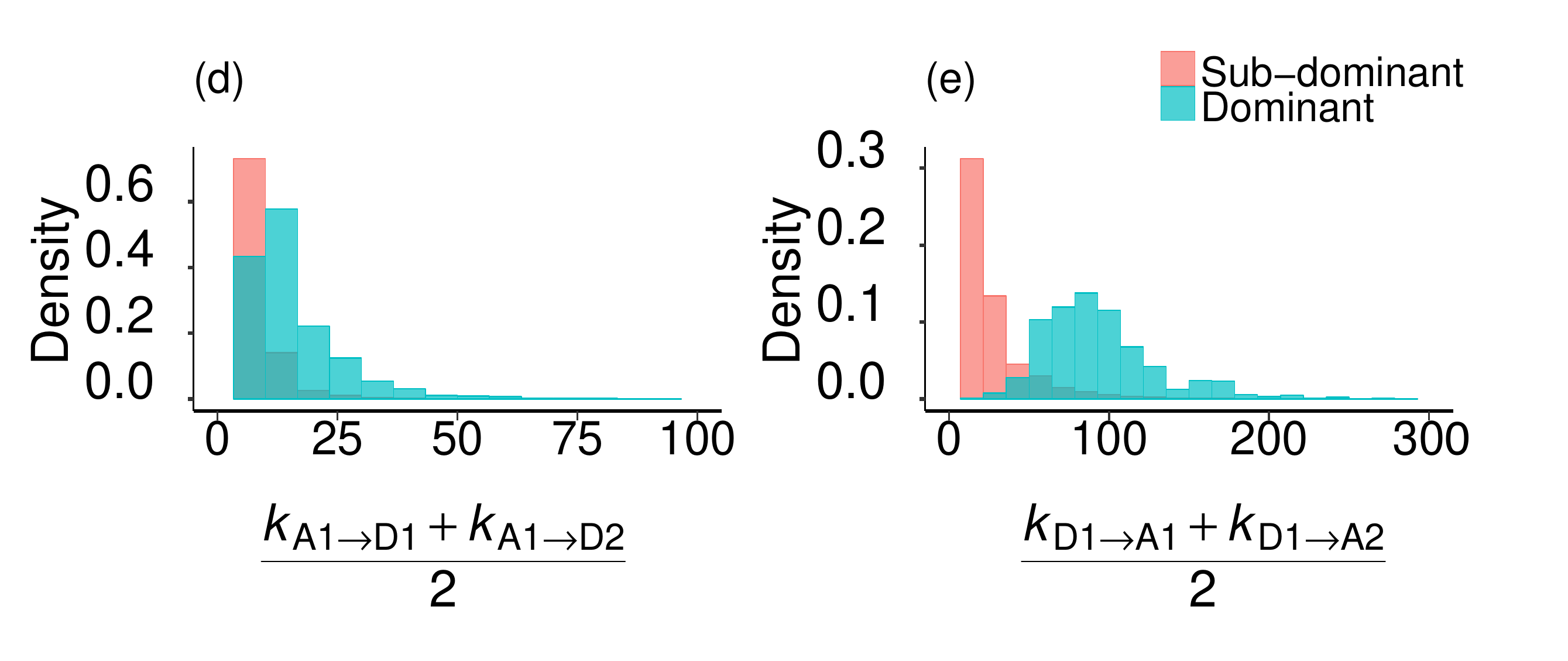}
\caption{(a) Deactivator essentiality and deactivator phases: on the \textit{y}-axis, +1 indicates only Deactivator-1 is essential, -1 indicates only Deactivator-2 is essential, while 0 indicates both are essential \cite{zullfikar_note}. We notice regions where Deactivator-1 remains essential and Deactivator-2 flip-flops in essentiality, which we designate as Deactivator-1 phase. In blue, +1 indicates Deactivator-1 phase and -1 indicates Deactivator-2 phase. (b) Activation Asymmetry and activator phases: in red, +1 indicates Activator-1 phase, -1 indicates Activator-2 phase. (c) Deactivation Asymmetry: both Activation and Deactivation Asymmetry correlate strongly with their respective phases. (d) Distributions of average $k_{\textrm{A} \rightarrow \textrm{D}}$ for $\textrm{A}_1$ dominant and subdominant. (e) Distributions of average $k_{\textrm{D} \rightarrow \textrm{A}}$ for $\textrm{D}_1$ dominant and subdominant. }
\label{2_Activation_Deactivation_Asym_TS}
\end{figure}

As a first step toward understanding the difference in these distributions, we examine the dynamics of evolution. To this end, it is helpful to introduce the concept of protein essentiality: we define a protein as being essential if its removal from the network causes oscillations to stop. As the network evolves, we observe periods of time when only Activator-1 is always essential, while Activator-2 flip-flops in essentiality, and periods where Activator-2 remains essential while Activator-1 flip-flops. The same is true for deactivators (see Fig. \ref{2_Activation_Deactivation_Asym_TS}a).  It was argued in \cite{PhysRevE.97.040401} that these long evolutionary periods reflect the division of sequence space into regions or phases separated by geometric bottlenecks (in sequence space). We define the ``phase'' associated with a protein in terms of  time periods during which it always stays essential, so that in Activator Phase 1, for example, Activator-1 remains essential while Activator-2 flip-flops in essentiality. 
Fig. \ref{2_Activation_Deactivation_Asym_TS} (b-c) show time that series plots of Activation and Deactivaction Asymmetry correlate well with the protein phases. Typically, the magnitude of a rate constant for a protein acting on a target is higher if the protein is in its associated phase.

To better understand the difference in asymmetry distributions for activators versus deactivators, we introduce the concepts of dominant and subdominant proteins: in Activator Phase 1,  we say Activator-1 is the dominant protein while Activator-2, which flip-flops in essentiality, is subdominant, with a similar definition for deactivators. Is there a relationship between whether a deactivator is dominant or subdominant and the strength of its associated chemical rate constants $k_{D \rightarrow A}$? For instance, in Fig. \ref{networkt500}, Deactivator-1 is dominant and its associated rate constants for suppressing the activators, $k_{D \rightarrow A}$, are also significantly stronger than those for the subdominant Deactivator-2. The distributions of $k_{\textrm{A} \rightarrow \textrm{D}}$ and $k_{\textrm{D} \rightarrow \textrm{A}}$ in dominant and subdominant phases depicted 
in Fig. \ref{2_Activation_Deactivation_Asym_TS} (d-e) provide support for the conjecture that the chemical rate constants,  $k_{\textrm{A} \rightarrow \textrm{D}}$ or $k_{\textrm{D} \rightarrow \textrm{A}}$, associated with the subdominant activator or deactivator are suppressed in comparison to the dominant activator or deactivator. Notice however, that while the distributions for the dominant and subdominant activators differ only modestly, the distributions for the dominant versus subdominant deactivators are strikingly different. 

Based on the results in Fig. \ref{2_Activation_Deactivation_Asym_TS}(d-e), we can now understand the bimodality in Deactivation Asymmetry depicted in Fig. 2 in terms of the pronounced suppression of $k_{\textrm{D} \rightarrow \textrm{A}}$ for the subdominant deactivator. Notice that this suppression arises naturally from neutral drift without any direct evolutionary selection pressure. To gain an intuitive understanding of this suppression, consider the following. 
An oscillatory cycle begins with low levels of activators and deactivators. Due to self-activation, the concentrations of activators in their active state start to rise, with the dominant activator typically leading the subdominant activator. The rising levels of activators cause the active deactivator concentrations to rise as well, with the dominant deactivator typically leading the subdominant deactivator. As the active level of the dominant deactivator rises, it starts to suppress both activators, leading to the peak and subsequent drop in active activator concentrations. This dynamics, necessary for sustained oscillations, is highly sensitive to the rate constants $k_{\textrm{D} \rightarrow \textrm{A}}$. Our hypothesis is that it is easier to generate sustained oscillations if the deactivation of the activators is strongly coupled to the active level of the leading (dominant) deactivator and only weakly coupled to the lagging (subdominant) deactivator.
To check this, we carried out the following test: we let a 2A-2D network evolve for 1000 accepted mutations and replaced $k_{\textrm{D}_1 \rightarrow \textrm{A}_1}$ and $k_{\textrm{D}_2 \rightarrow \textrm{A}_1}$ by their average value (and the same for $\textrm{A}_2$) and determined if the network continued to oscillate upon making this change. These results were then compared to the case where we replaced $k_{\textrm{A}_1 \rightarrow \textrm{D}_1}$ and $k_{\textrm{A}_2 \rightarrow \textrm{D}_1}$ by their average value (and the same for $\textrm{D}_2$). For the deactivators acting on activators, only $19.1 \%$ of these average-value substitutions resulted in oscillations. On the other hand, for activators acting on deactivators, 
$65.6 \%$ of the average-value substitutions resulted in oscillators.  These results imply that our network is able to yield oscillations relatively easily when the rates $k_{\textrm{A}_1 \rightarrow \textrm{D}_1}$ and $k_{\textrm{A}_2 \rightarrow \textrm{D}_1}$ are comparable (similarly true for activation of $\textrm{D}_2$), but has difficulty producing oscillations when the rates $k_{\textrm{D}_1 \rightarrow \textrm{A}_1}$ and $k_{\textrm{D}_2 \rightarrow \textrm{A}_1}$ are comparable (similarly true for deactivation of $\textrm{A}_2$). However, this test reveals only part of the picture, as replacement by the average reduces the effect of  the larger rate constant and increases the effect of the smaller. Is it one or both of these changes that matter? 

We answered this question by carrying out the same test as before but this time replacing the larger and the smaller $k_{\textrm{D} \rightarrow \textrm{A}}$ by their average separately, and determining whether the network continued to oscillate. For activators acting on deactivators, replacing the smaller $k_{\textrm{A} \rightarrow \textrm{D}}$ by the average resulted in $61.0 \%$ oscillators, and replacing the larger $k_{\textrm{A} \rightarrow \textrm{D}}$ by the average resulted in $38.4 \%$ oscillators. For deactivators acting on activators, replacing the smaller $k_{\textrm{D} \rightarrow \textrm{A}}$ by the average resulted in $21.6 \%$ oscillators, and replacing the larger $k_{\textrm{D} \rightarrow \textrm{A}}$ by the average resulted in only $2.50 \%$ oscillators. These results imply that either lowering $k_{\textrm{D} \rightarrow \textrm{A}}$ for the dominant deactivator or increasing $k_{\textrm{D} \rightarrow \textrm{A}}$ for the subdominant deactivator results in far fewer oscillators, with the effect being particularly pronounced in the case of the weakening dominant deactivator. By contrast, changing $k_{\textrm{A} \rightarrow \textrm{D}}$ for either the dominant or subdominant activator has a less pronounced effect on oscillations.
\section{Synergy and Competition}
\begin{figure}[h!]
\includegraphics[scale=0.3]{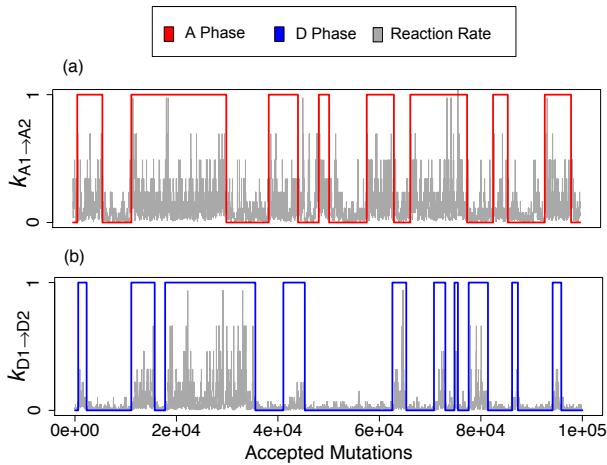}
\caption{Synergistic activation between activators versus competitive deactivation between deactivators. The activator (deactivator) phase is shown in red (blue), while $k_{\textrm{A}_1 \rightarrow \textrm{A}_2}$  ($k_{\textrm{D}_1 \rightarrow \textrm{D}_2}$) is shown in gray. +1 indicates Activator-1 (Deactivator-1)  phase, while 0 indicates Activator-2 (Deactivator-2) phase. (a) The activation rate of an activator by the other activator, regardless of dominance or subdominance, does not correlate well with phase. (b) The deactivation rate of the subdominant deactivator is strongly correlated with the phase of the dominant deactivator.}
\label{symbiotic_parastiic}
\end{figure}

  This brings us to a further distinction in the relationship between activators versus that between deactivators. Activator-1, for example, can activate Activator-2, which in turn further activates Activator-1, so that they act synergistically and effectively increase the degree of cooperativity for autoactivation. This is different for deactivators: Deactivator-1 suppresses Deactivator-2, and vice versa, so that deactivators act competitively, with the dominant deactivator suppressing the subdominant deactivator. This important distinction is consistent with our observation that activation of one activator by the other, regardless of its dominance or subdominance, does not correlate well with protein phase, implying synergistic activation of both activators for all evolutionary periods (e.g., see Fig. \ref{symbiotic_parastiic}a). On the other hand, the deactivation of one deactivator by the other correlates strongly with deactivator phase; this behavior implies that when a deactivator becomes dominant the rate at which it \textit{deactivates} the subdominant deactivator typically increases; an example of this behavior can be seen in Fig. \ref{symbiotic_parastiic}b.  This idea of synergistic  versus competitive interactions is further borne out by the observation that the dominant activator has a higher autoactivation rate, whereas a subdominant deactivator has a higher auto-deactivation rate (see Fig. \ref{Supplemental_figure_1}, \cite{suppm_note}).  We can quantify this difference by carrying out another rate replacement test: we let a 2A-2D network evolve for 1000 accepted mutations but now replace the smaller of $k_{\textrm{A}_1 \rightarrow \textrm{A}_2}$ and $k_{\textrm{A}_2 \rightarrow \textrm{A}_1}$ by the larger value and determine if the network continued to oscillate. These results are then compared to replacing the smaller of $k_{\textrm{D}_1 \rightarrow \textrm{D}_2}$ and $k_{\textrm{D}_2 \rightarrow \textrm{D}_1}$. Based on the above analysis, we expect oscillations to be more likely to persist in the case of activators than deactivators, since activators are more likely to cooperate. Indeed, we find that 42.2\% of tests resulted in oscillators for activator rate replacement while only 22.6\% of the tests resulted in oscillators in the case of deactivator rate replacement. 
 
\section{Asymmetry in Cooperativity}
We have discussed that the subdominant activator serves to increase the rate and also the cooperativity of effective autoactivation of the dominant activator. Is there an asymmetry in the role of cooperativity in autoactivation versus auto-deactivation in the context of oscillations? To address this question, we study a 1A-1D network and perform a stability analysis about its fixed points. To determine stability of a steady state, one must know the eigenvalues of the Jacobian matrix \textbf{J} of the system evaluated at the steady state~\cite{Tysonbib}. However, for the special case of the two-node network, it is sufficient just to know the trace and determinant of the Jacobian. If the determinant $\text{det}(\textbf{J}) > 0$ and the trace $\text{tr}(\textbf{J})<0$, then the steady state is stable and we do not expect limit-cycle oscillations. We will show that the trace becomes negative and the determinant becomes positive for any value of exponent of autoactivation that is less than a critical value. For notational convenience, in the following discussion we will use $A^*$ and $D^*$ to represent the active concentrations, and $A$ and $D$ to represent the inactive concentrations of the activator and deactivator, respectively. For a 1A-1D network, the dynamics of the active fractions of concentrations are given by:
\begin{equation}
\frac{dA^*}{dt} = k_{11} A{^*}^a A - k_{21}{D^*}^2 {A^*} + \alpha{A} - \beta{A^*},
 \label{dkstardt}
\end{equation}
\begin{equation}
\frac{dD^*}{dt} = k_{12} {A^*}^2 {D} - k_{22}{D^*}^b{D^*} + \alpha{D} - \beta{D^*} ,
 \label{dkstardtD}
\end{equation}
\noindent where we have left the exponents of the autoactivation and auto-deactivation as variables $a$ and $b$ respectively. We denote the fixed points of the system by $A^{*}_{0}$ and $D^{*}_{0}$; they are obtained by setting $d A^{*}/dt$ and $d D^{*}/ dt$ to zero in Eqs. \ref{dkstardt} and \ref{dkstardtD} and solving for the active fractions of concentration. The Jacobian matrix of the system at the fixed point is
\[
J=
  \begin{bmatrix}
    a_{11} & a_{21} \\
    a_{12} & a_{22}
  \end{bmatrix}
\]
with components defined as follows:
\begin{itemize}
\item{$a_{11} = a k_{11} (1-A^{*}_{0}) {A^{*}_{0}}^{(a-1)} - k_{11} {A^{*}_{0}}^a - k_{21}{D^{*}_{0}}^2 -\alpha -\beta $}
\item{$a_{12} = -2 k_{21} A^{*}_{0}D^{*}_{0}$}, ( \textless \hspace{0.1 cm} 0)
\item{$a_{21} =  2 k_{12} A^{*}_{0}{D}_{0}$}, ( \textgreater \hspace{0.1 cm} 0)
\item{$a_{22} = -k_{12} {A^{*}_{0}}^2 -(b+1)k_{22}{D^{*}_{0}}^b - \alpha - \beta$}, (\textless \hspace{0.1 cm} 0)
\end{itemize}
\noindent Inspection shows that $a_{12}$ is always negative, $a_{21}$ is always positive or equal to 0, and $a_{22}$ is always negative.  $a_{11}$ can be positive or negative depending on the value of $a$. If $a_{11} \le 0 $,  $\text{det}(\textbf{J}) > 0$ and $\text{tr}(\textbf{J})<0$, implying the 
absence of oscillations. It is thus necessary for $a_{11} > 0$ for the system to oscillate. 
To determine the role of cooperativity for producing oscillations, we rewrite Eq. \ref{dkstardt} as
\begin{equation}
k_{11} {A^{*}_{0}}^a(1-A^{*}_{0}) = k_{21}{D^{*}_{0}}^2A^{*}_{0} + \alpha{A}_{0} - \beta A^{*}_{0} .
 \label{dkstardt0}
\end{equation}
\noindent Dividing Eq. \ref{dkstardt0} by $A^{*}_{0}$ and plugging into the expression for $a_{11}$, we find
\begin{equation}
a_{11} =  - \frac{a\alpha}{ A^{*}_{0}}  -k_{11}{{A^{*}_{0}}}^a + (a-1)(k_{21}{D^{*}_0}^2)+(a-1)(\alpha+\beta).
 \label{a_11}
\end{equation}
Note that at $a=1$, corresponding to the absence of cooperativity in autoactivation, we have
\begin{equation}
a_{11} =  - \frac{a\alpha}{A^{*}_{0}}  -k_{11}{A^{*}_{0}}^a <0,
\end{equation}
\noindent so that $\textrm{det(\textbf{J})} > 0$. Consequently, there can be no oscillations for $a = 1$. We find also that there is no such constraint imposed by cooperativity in auto-deactivation, e.g. we verified numerically that both $a_{11} $ and $a_{11} + a_{22}$ can be greater than zero for $b=1$. This analysis highlights the asymmetry in the role of cooperativity in autoactivation versus auto-deactivation for producing oscillations.

\section{Conclusion}

In this paper, we employed a sequence-based mutational algorithm to study the evolution of oscillatory protein networks. Beginning from a core module of an activating and deactivating protein that are subsequently duplicated, we found that deactivators possess a high degree of Deactivation Asymmetry while activators do not display any such Activation Asymmetry (see Fig. \ref{Asymmetry_Statistics}). We can understand this asymmetry by the synergistic roles of activating proteins and competitive roles of deactivating proteins: when an activator becomes subdominant, the dominant activator in the network works to increase the former's activity. On the other hand, the dominant deactivator suppresses the subdominant deactivator. Finally, we showed that cooperativity is required only in autoactivation and not in auto-deactivation to generate oscillations.

Our theoretical results imply strong asymmetries in activator versus deactivator essentiality and function. We believe that more experimental work will further reveal the impact of such asymmetric behavior in protein-protein networks. Indeed, recent studies have already found these asymmetries in `omics' datasets of kinases and phosphatases \cite{smoly2017asymmetrically}. An interesting future direction will be to explore these ideas quantitatively using a partial information decomposition, as in \cite{PhysRevE.91.052802}. Another future direction might also be to extend our model to systems other than oscillatory networks, such as signaling networks, to theoretically investigate asymmetries in the organization and dynamics of activators and deactivators. Finally, it would be interesting to see how the asymmetric roles of activators and deactivators extend to networks where the number of nodes is not fixed and could change with evolution.

We acknowledge helpful discussions with Md. Zulfikar Ali. This work was supported in part by the National Science Foundation, Grant No. PHY-1607612, and the National Institutes of Health, Grant No. R01 GM082938.

\bibliographystyle{apsrev4-1}
\bibliography{second_paper}

\begin{thebibliography}{17}
\bibitem [{zul()}]{zullfikar_note}%
  \BibitemOpen
  \href@noop {} {}\bibinfo {note} {Since we find that states where both
  activators are individually inessential are very rare, approximately
  $0.001\%$ of the total number of oscillatory states, we ignore such states
  for the purposes of the figure.}\BibitemShut {Stop}%
\end{thebibliography}

\pagebreak
\widetext
\begin{center}
\textbf{\large Supplemental Materials: Asymmetry between Activators and Deactivators in Functional Protein Networks}
\end{center}
\setcounter{equation}{0}
\setcounter{figure}{0}
\setcounter{table}{0}
\setcounter{page}{1}
\makeatletter
\renewcommand{\theequation}{S\arabic{equation}}
\renewcommand{\thefigure}{S\arabic{figure}}
\renewcommand{\bibnumfmt}[1]{[S#1]}
\renewcommand{\citenumfont}[1]{S#1}

\begin{figure}[h]
\includegraphics[scale=0.5]{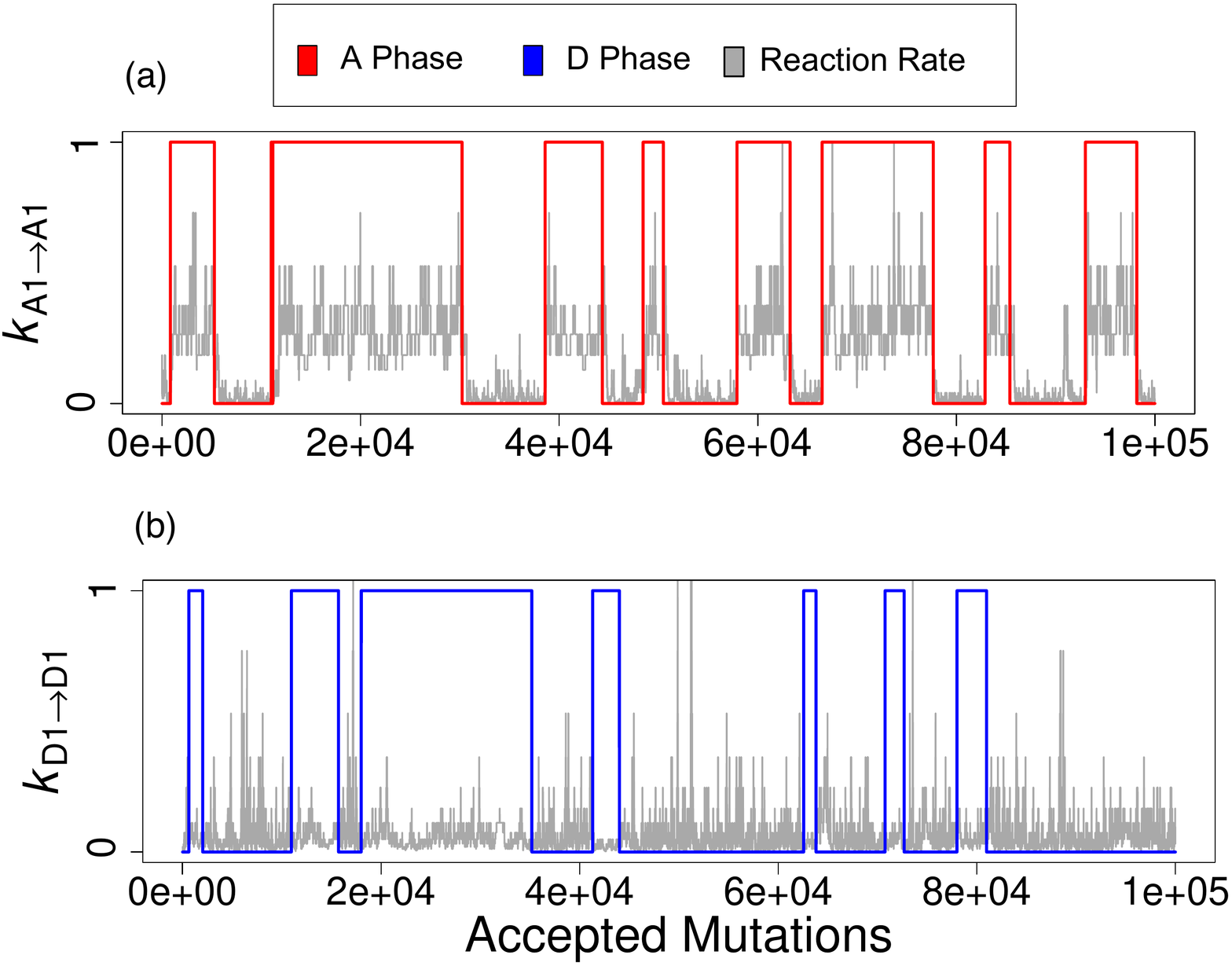}
\caption{(a) Autoactivation and Activator phase.  In red, 1 indicates Activator-1 phase and 0 indicates Activator-2 phase. (b) Auto-deactivation and Deactivator phase. In blue, 1 indicates Deactivator-1 phase and 0 indicates Deactivator-2 phase. The dominant activator has a higher autoactivation rate, whereas a subdominant deactivator has a higher auto-deactivation rate.}
\label{Supplemental_figure_1}
\end{figure}


\end{document}